\documentclass[a4paper,11pt]{article}
\usepackage{amssymb}
\usepackage{extarrows}

\pdfoutput=1 

\usepackage{jinstpub}

\title{\boldmath Time Projection Chamber as Inner Tracker for Super Charm-Tau Factory at BINP}

\author[a]{V. K. Vadakeppattu}
\author[a,b]{A. V. Sokolov}
\author[a,b]{L. I. Shekhtman}
\author[a,b]{T. V. Maltsev}

\affiliation[a]{Novosibirsk State University}
\affiliation[b]{Budker Institute of Nuclear Physics}

\emailAdd{vijayanandkv.anand@gmail.com}

\abstract {At present time Budker INP is developing a Super Charm-Tau factory project, which consists of a high-luminosity collider with the luminosity of $10^{35}$\,cm$^{-2}$s$^{-1}$ and a universal magnetic detector. The tracking system of the detector will comprise of an Inner Tracker (IT) and a Drift Chamber (DC). One of the options for IT is Time Projection Chamber (TPC). The advantages of the TPC are high spatial resolution and particle identification capabilities by registration of dE/dx losses. However, using a Time Projection Chamber implies serious challenges. For example, the TPC have to simultaneously deal with tracks from several thousands  events and maintain the enormous data rate. This work describes the results of the Monte-Carlo studies of the transport characteristics	in various gas mixtures proposed for TPC. Besides of this, the simulation of the ion back flow and its effect on spatial resolution will be reported.}

\proceeding{Instrumentation for Colliding Beam Physics\\
24 to 28 February, 2020\\
Budker Institute of Nuclear Physics, Novosibirsk,Russia}

\begin{document}
\maketitle
\flushbottom

\section{Introduction}

\begin{figure}[tbph]
	\centering
	\includegraphics[width=0.4\textwidth]{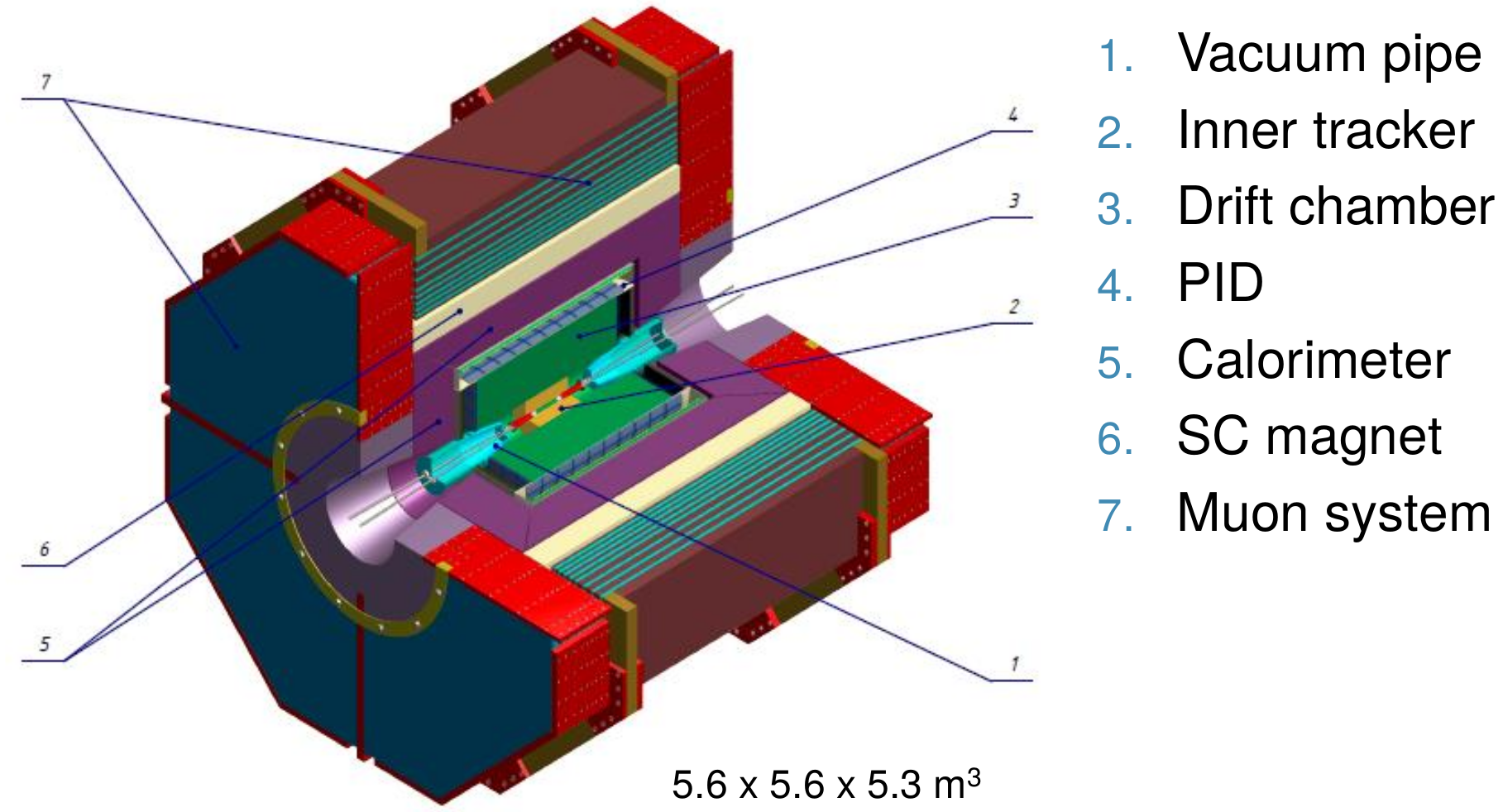}
	\caption{\label{fig:detector} Detector for the Super Charm-Tau factory in Novosibirsk.}
\end{figure}

Objective of the Super Charm-Tau factory at Novosibirsk is to study the rare decays of D-meson, $\tau$-lepton, unobserved $\tau$ decays etc. Collider has an energy range of 2-7\,GeV in the center of mass system (CMS) and luminosity $10^{35}$\,cm$^{-2}$s$^{-1}$~\cite{a}. The broad physics program requires the development of a universal magnetic detector (Fig.~\ref{fig:detector}). The tracking system of the proposed Super Charm-Tau factory Detector (SCTD) consists of the Inner Tracker (IT) and the Drift Chamber(DC). The inner tracker is placed between the beam pipe and the drift chamber. It provides the detection solid angle up to 98\,\%. The inner tracker is a cylinder with a length of 60\,cm, inner diameter is 5\,cm,
outer diameter is 40\,cm. 

As it is placed near to the beam pipe, the inner tracker has to handle high particle flux. The main aim of the inner tracker is to detect secondary vertices from decays of short lived particles, like $K_{s}$ and $\Lambda$, and to help drift chamber to measure the momentum of charged particles owing to the increased lever arm. This is especially refers for soft charged particle tracks. Information from the IT can either be processed with that of drift chamber or alone. The cut off value for transverse momentum of the pions, penetrating into the IT is about 60 MeV/c~\cite{b} due to the material of the vacuum pipe. The candidates for the inner tracker are the Time Projection Chamber (TPC), the Silicon strip detector and cylindrical MPGD detector. Material budget for these options are shown in table~\ref{table:budget}. This paper describes results of Monte Carlo studies on transport properties for various gas mixtures proposed for TPC.

\begin{table}[h]
  \caption{Material budget for the different inner tracker options}
  \centering
  \begin{tabular}{ | l | c | c |}
  \hline
   Option/Subsystem & Material 												 & Thickness (in X$_0$)      \\ \hline
   Vacuum pipe 		&1\,mm Be + 0.6\,mm Paraffin + 0.6\,mm Be                &  0.9\,\%  \\ \hline
   TPC              & 2 $\times$ (1\,mm G10 + 0.1\,mm Teflon +15$\mu$\, Cu ) &  1.5\,\%  \\ \hline
   C-MPGD       	& 4 $\times$ (0.25\,mm Kapton +40$\mu$\,Cu )             &  1.4\,\%  \\ \hline
   Si-strip         & 4 $\times$ (0.32\,mm Si +0.4\,mm carbon fiber )        &  1.4\,\%  \\ \hline
  \end{tabular}
  \label{table:budget}
\end{table}

\section{Time Projection Chamber as Inner Tracker}
Time Projection Chamber is considered as a good option for the inner tracker detector in SCTD as it is an ideal device for tracking of charged particles in three dimensional space by fulfilling all the above requirements. Also, by allowing to measure dE/dx, it is possible to identify particle masses using TPC. Several experiments such as ALICE~\cite{c}, ILD~\cite{d} etc are considering TPC as tracker.

\begin{figure}[tbph]
	\centering
	\includegraphics[width=0.4\textwidth]{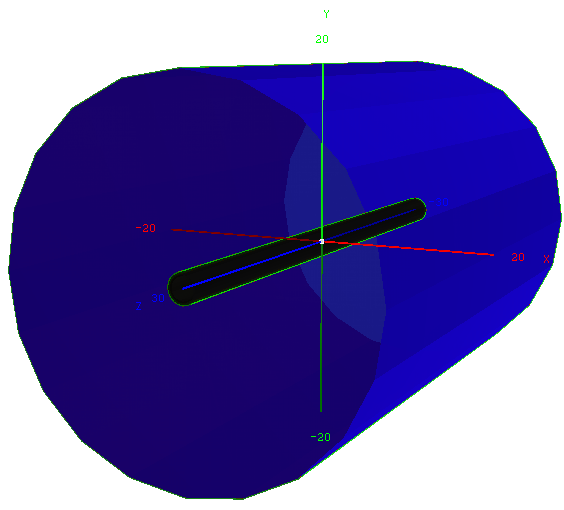}
	\caption{\label{Figure:2} 3-D view of TPC from detector simulation}
\end{figure}

\par TPC is a large cylindrical volume filled with gas, surrounding interaction point by covering almost full solid angle. TPC for c-tau detector have length of 600\,mm, inner and outer diameter of 50\,mm and 400\,mm. A uniform electric field of 500\,V/cm is applied along Z-direction and magnetic field of 1.5\,T is applied parallel to electric field. Studies of two TPC options with triple GEM and $\mu$-RWELL readouts are going on.

\section{Choice of the drift gas mixtures}
Although gas detectors have been used in experimental physics for several decades, there is no specific recipe for choosing the gas mixture. When choosing a mixture for a particular detector, the following considerations should be taken into account: ionization density, radiation thickness, electron drift velocity and diffusion, ion drift velocity, quenching properties and chemical stability of the gas mixture components.One of the most common are gas mixtures based on Argon or Neon (see Table~\ref{table:gases}). Helium is often used when large radiation length is needed, but for the inner tracker, the contribution of gas mixture is small compared with the material which the TPC is made of (see Table~\ref{table:budget}). Indeed even for Ar based mixtures radiation thickness is less than 0.1\,\% of X$_0$.

\begin{table}[h]
	\caption{Properties of gases commonly used in TPCs at normal temperature and pressure. Density $\rho$, radiation length X$_0$, total number of electron-ion pairs for MIPs N$ _t $.}
	\centering
	\begin{tabular}{ | l | c | c | c | c |}
		\hline
		Gas		 	& $\rho$ & X$_0$      & X$_0$ &  N$ _t $    \\ 
	        	 	& [g/l]  & [g/cm$^2$] & [m]   &  [1/cm]     \\ \hline
		He  	 	& 0.1785 & 94.32      & 5280  &  8          \\ 
		Ne  	 	& 0.8999 & 28.94      & 322   &  40         \\ 
		Ar  	 	& 1.784  & 19.55      & 110   &  97         \\ \hline
		CH$_4$   	& 0.717  & 46.22      & 645   &  54         \\ 
		CO$_2$   	& 1.977  & 36.2       & 183   &  100        \\ 
		C$_2$H$_6$ 	& 1.977  & 45.47      & 335   &  112        \\ 
		CF$_4$   	& 3.93   & 36 	      & 90    &  120        \\ \hline
	\end{tabular}
	\label{table:gases}
\end{table}

\subsection{Field distortion due to space charge }
\begin{figure}
	\includegraphics[width=0.45\textwidth]{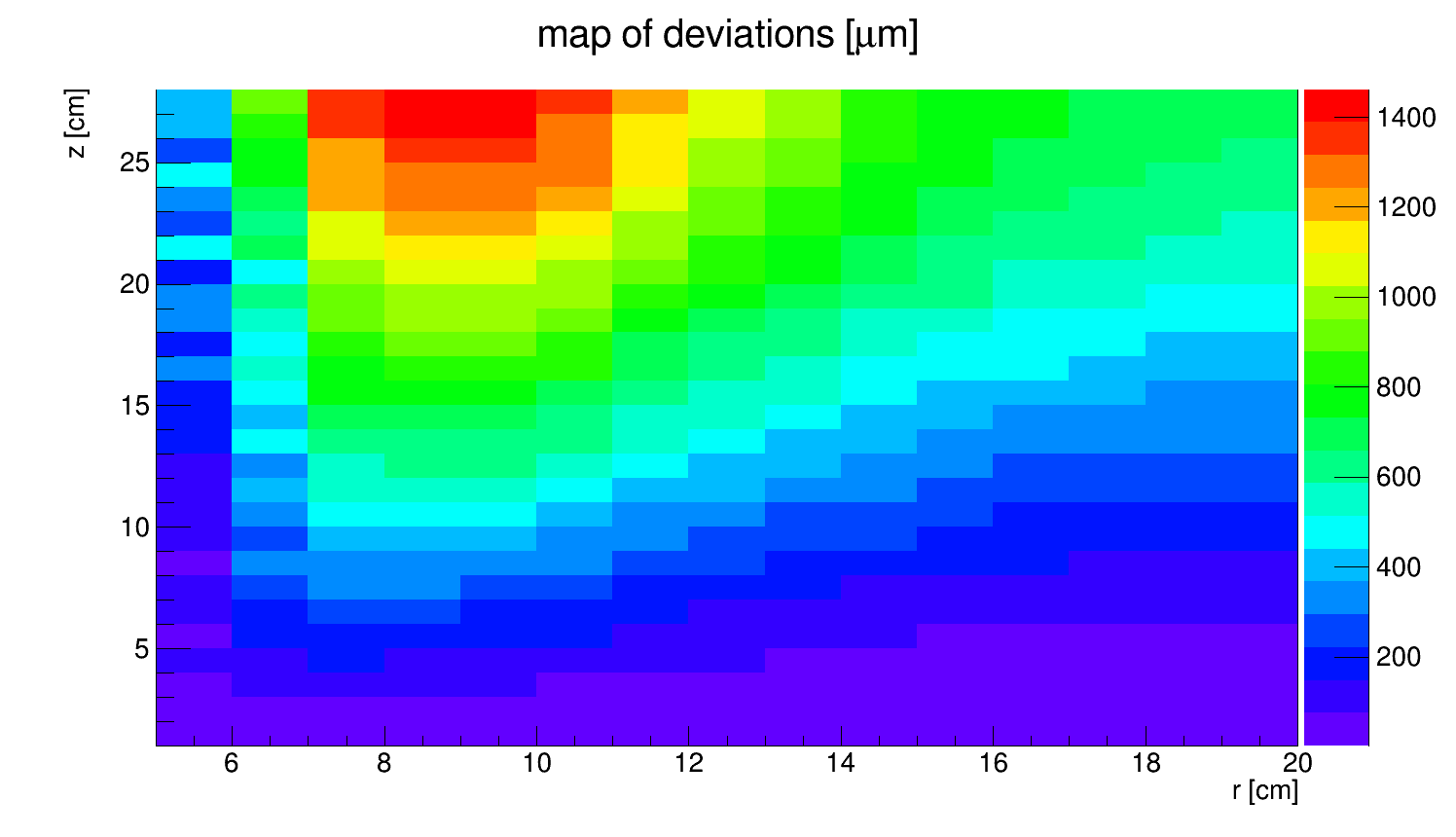}
	\qquad
	\includegraphics[width=0.45\textwidth]{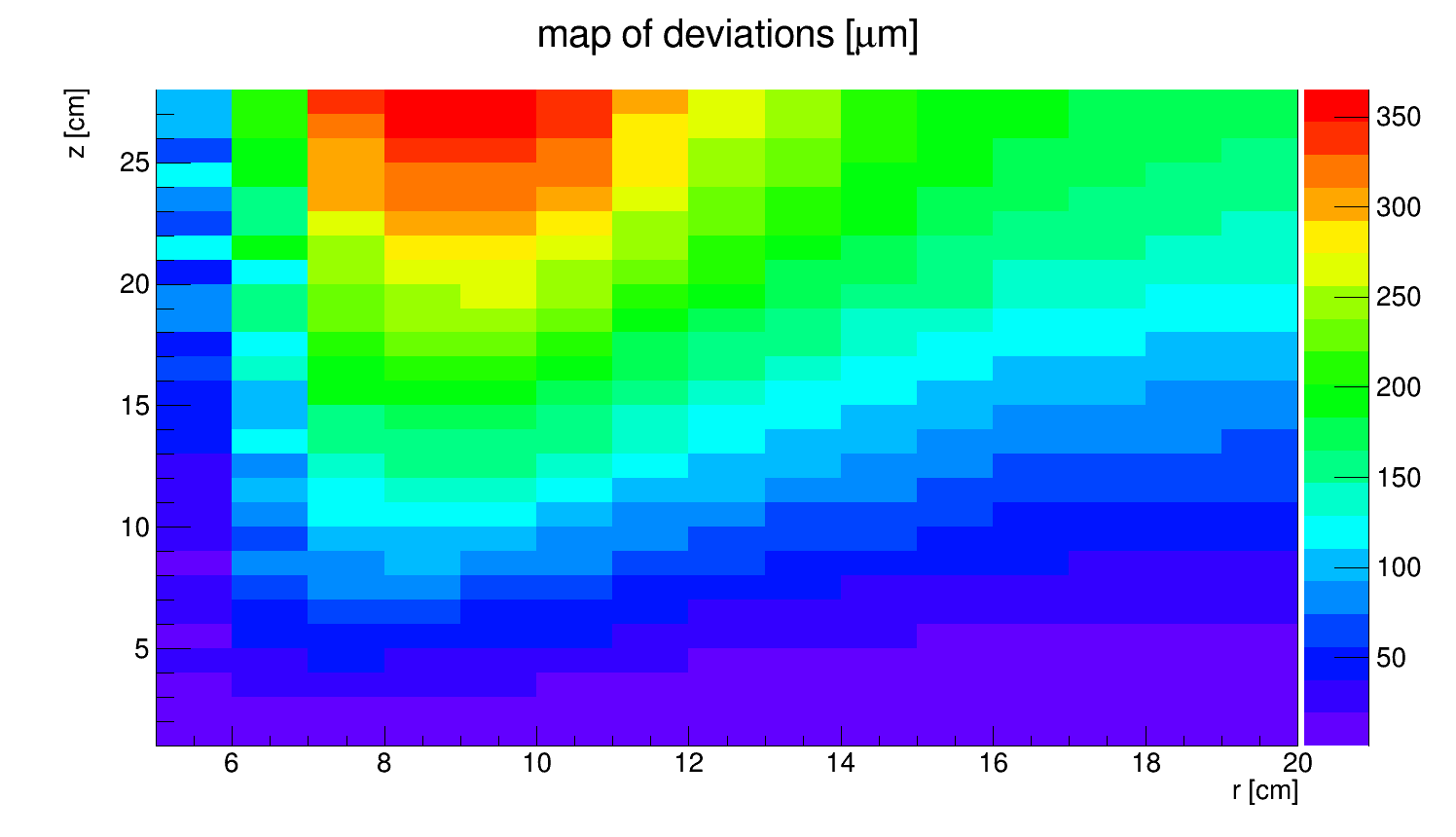}
	\caption{\label{fig:distortion} The deviation of the drift lines distortion from straight trajectories for full drift length (30\,cm) in Ar (left) and Ne (right) based gas mixtures. r- and z-axes correspond to directions along the TPC radius and the beam, respectively. Deviations are given in microns. }
	
\end{figure}

One of the main limitations for the TPC application in continuous read-out mode is the electric field distortion of a uniform electric field due to a space charge of ions, appearing in the Ion Back Flow (IBF) process. IBF appears in the triple-GEM at the end-cap of TPC, ions from the avalanche in GEMs move to TPC volume. With regular triple-GEM cascade, the value of IBF depends on the fields inside GEM holes and in the drift and transfer gaps, and can be easily obtained around 10\%. With special efforts, using 4-GEM cascade with different hole diameters and with different voltages across GEMs and transfer fields, the value of 1\% of IBF can be obtained~\cite{c}. While moving essentially slower (about 1000 times) than electrons, ions stay in TPC volume longer and produce their own  electric field. This electric field distorts straight external electric field lines in TPC. The results of physics background simulation~\cite{f} were used for space charge density estimation in TPC. The electric field distortion calculation was based on the obtained values of space charge density with IBF 1\% and gain 10000. The ions space charge density was about $10^7 |e|/cm^3$ (|e| is an absolute value of electron charge) near beam pipe inside TPC and it decreased strongly up to $10^4 |e|/cm^3$ near the outer wall of TPC. In Fig.~\ref{fig:distortion}, the field line deviation map from straight field lines is shown for drift length 30 cm in Ar and Ne based gas mixtures. As charge particles move along field lines, the distortion of field lines is the distortion of the particle trajectories. To evaluate distortion of the field lines due to the effect of space charge the COMSOL simulation package have been used~\cite{g}. Despite the fact that the space charge in the Ar based mixtures is four times larger than in the Ne-based (due to twice higher ionization and two times lower ions drift velocity compared to Neon), the deviations of drift lines in Ar ($\sim1.5\,$mm) are comparable to the transversal diffusion and can be compensated by offline analysis. It makes Ar-based gas mixtures favorable as working medium in the TPC due to low cost and insensitivity to impurities.
\begin{figure}
	\includegraphics[width=0.45\textwidth]{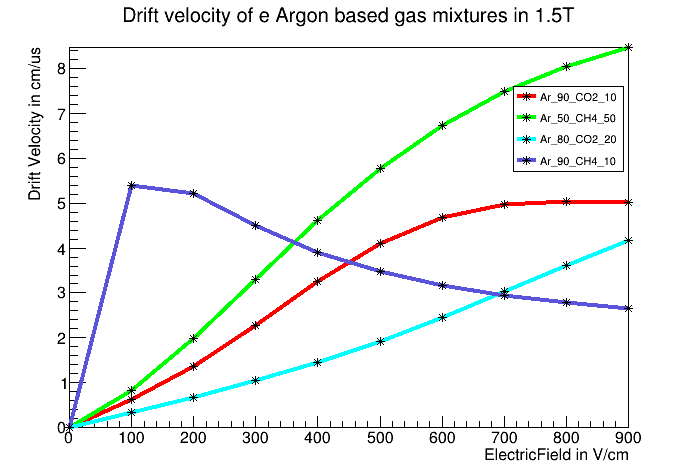}
	\qquad
	\includegraphics[width=0.45\textwidth]{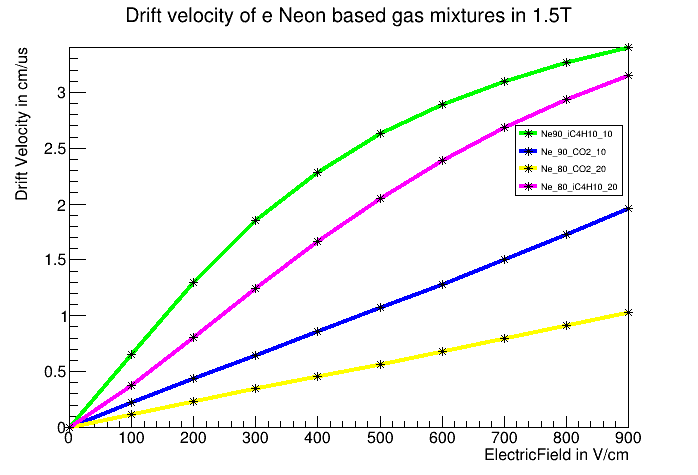}
	\caption{\label{fig:velocity} Drift velocity of electron in  argon and neon-based gas mixtures with magnetic field 1.5T applied parallel to electric field.}
\end{figure}
\begin{figure}
	\includegraphics[width=0.45\textwidth,trim=10 10 0 0,clip]{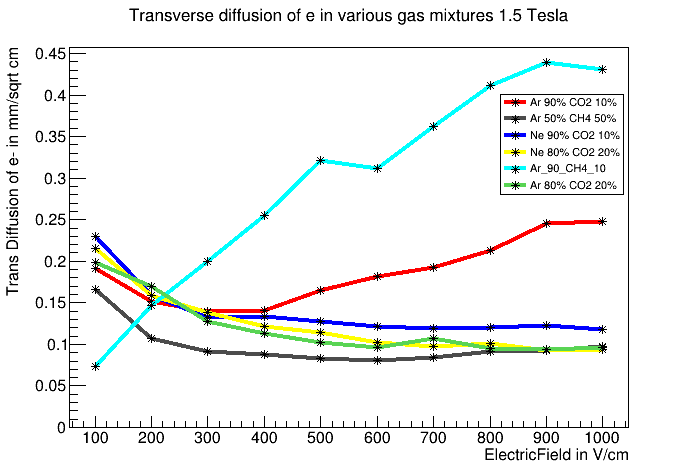}
	\qquad
	\includegraphics[width=0.45\textwidth,origin=c,angle=0]{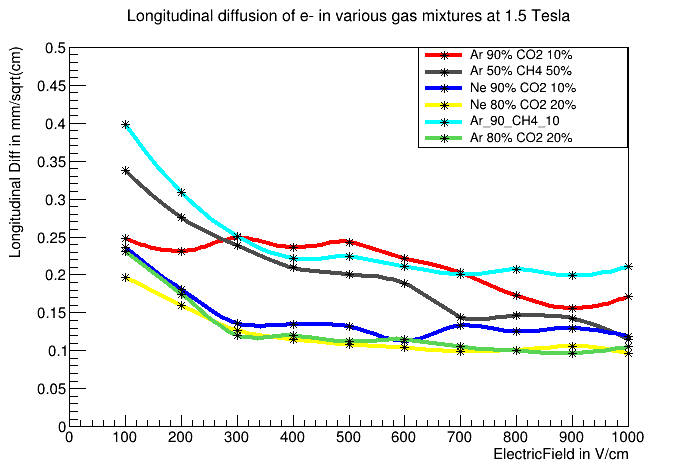}
	\caption{\label{fig:diff} Transverse and longitudinal diffusions as a function of electric field with magnetic field of 1.5T applied parallel to electric field.}
	
\end{figure}

\subsection{Diffusion and drift velocity}
Transport properties of electrons in various gas mixtures have been studied using Garfield++ software package~\cite{h}. The drift velocity in Argon based gas mixtures is significantly higher compared to neon, due to Ramsauer effect~\cite{i}, as can be seen in Fig.~\ref{fig:velocity}. High drift velocity means less events overlap in TPC, which simplifies event reconstruction. For example for Ar/CO$_2$ (90/10) gas mixture at an electric field 500\,V/cm, tracks from about 6000 particles will simultaneously present in the TPC (that corresponds 1500 bunch crossings). For comparison, with a gas mixture of Ne/CO$_2$ (90/10) at 500\,V/cm there will present about 24000 tracks in TPC volume. In this regard, a mixture of Ar/CH$_4$ (90/10) is of great interest, having  at an electric field of only 125\,V/cm the drift velocity of 5\,cm/$\mu$s. That makes the design of the field cage much simpler. Another gas mixture Ar/CH$_4$ (50/50), is interesting because it allows to maximize drift velocity, thereby minimizing the events overlap. Price to pay for this is a significant complication of the design of the field cage, since this mixture requires a high electric field strength up to 1000\,V/cm.
Fig.~\ref{fig:diff} shows the transverse and longitudinal diffusion of electrons in gas mixtures. It is clear from the graph that in 125\,V/cm Ar/CH$_4$ (90/10) gas mixture has the smallest transverse diffusion among all others, that makes it even more attractive to use. At 500\,V/cm Ar/CO$_2$ (90/10) has the slightly higher transverse diffusion than Ne/CO$_2$ with the same proportion. However, Ar/CO$_2$ (90/10) has the highest longitudinal diffusion followed by Ar/CH$_4$ (90/10) at 500\,V/cm. The values of transverse and longitudinal diffusion determines charge cluster size in R-Phi and Z directions correspondingly. The value of spatial resolution, however, might be much smaller and depends on particular size of readout pads, electronic noise and the method of calculation of center of gravity of induced charge.

\section{Conclusions}

  The main goal of this work is to propose a gas mixture which can be used for the continuous operation of TPC with MPGD readout. Simulation of the drift lines distortion due to the spatial charge shows that it is possible to use an argon based gas mixture for the TPC operation. Such mixtures are preferred in terms of ionization drift time and electric field strength. However, further study of the achievable spatial resolution for the selected gas mixtures is necessary.


\begin{thebibliography}{99}
\bibitem{a}
\emph{Super Charm-Tau Factory, https://ctd.inp.nsk.su/c-tau/}

\bibitem{b}
{T.V. Maltsev, L.I. Shekhtman, A.V. Sokolov and V.K. Vadakeppattu.} 
\emph{Simulation of different options of the Inner Tracker for	Novosibirsk Super Charm-Tau Factory Detector.}

{EPJ Web of Conferences 212, 01009 (2019)} 


\bibitem{c}
{ ALICE Collaboration, \emph{Technical Design Report for the Upgrade of the
	ALICE Time Projection Chamber}, ALICE-TDR-016, CERN-LHCC2013-020, March 3, 2014.}

\bibitem{d}
{R. Diener}
\emph{Development of a TPC for an ILC Detector}
{Science Direct 2012, physics procedia 37}

	
\bibitem{f}
{L. Shekhtman, F. Ignatov and V. Tayursky,  EPJ Web of Conferences 212, 01009 (2019).} 
\emph{Simulation of physics background in Super c-tau factory detector.}

{EPJ Web of Conferences 212, 01009 (2019)} 

\bibitem{g}
\emph{COMSOL Multiphysics, http://comsol.com}

\bibitem{h}
\emph{Garfield++, https://garfieldpp.web.cern.ch/garfieldpp/}

\bibitem{i}
{Sobhani H., Hassanabadi H. and Chung W.S. \emph{Observations of the Ramsauer–Townsend effect in quaternionic quantum mechanics.} Eur. Phys. J. C 77, 425 (2017).} 



\end{thebibliography}
\end{document}